# Ultrafast Demagnetization in a Ferrimagnet under Electromagnetic Field Funneling

Kshiti Mishra[a], Agne Ciuciulkaite[b], Mario Zapata-Herrera[c], Paolo Vavassori[c,d], Vassilios Kapaklis[b], Theo Rasing[a], Alexandre Dmitriev[e,*], Alexey Kimel[a], and Andrei Kirilyuk[a,f,*]

[a]Radboud University, Institute for Molecules and Materials, Heyendaalseweg 135, 6525 AJ Nijmegen, The Netherlands

[b]Department of Physics and Astronomy, Uppsala University, Box 516, SE-75120 Uppsala, Sweden

[c]CIC nanoGUNE BRTA, E-20018 Donostia-San Sebastian, Spain

[d]IKERBASQUE, Basque Foundation for Science, E-48009, Bilbao, Spain

[e]Department of Physics, University of Gothenburg, SE-412 96 Göteborg, Sweden

[f]FELIX Laboratory, Radboud University, Toernooiveld 7, 6525 ED Nijmegen, The Netherlands

Email: alexd@physics.gu.se, andrei.kirilyuk@ru.nl

**ABSTRACT**

The quest to improve density, speed and energy efficiency of magnetic memory storage has led to exploration of new ways of optically manipulating magnetism at the ultrafast time scale, in particular in ferrimagnetic alloys. While all-optical magnetization switching is well-established on the femtosecond timescale, lateral nanoscale confinement and thus potential significant reduction of the size of the magnetic element remains an outstanding challenge. Here we employ resonant electromagnetic energy-funneling plasmon nanoantennas to influence the demagnetization dynamics of a ferrimagnetic TbCo alloy thin film. We demonstrate how Ag nanoring-shaped antennas under resonant optical femtosecond pumping reduce the overall magneto-optical response due to demagnetization in the underlying films up to three times compared to non-resonant illumination. We attribute such substantial reduction to the nanoscale confinement of the demagnetization process. This is qualitatively supported by the electromagnetic simulations that strongly evidence the optical energy-funneling to the nanoscale from the nanoantennas into the ferrimagnetic film. This is the first and defining step for reaching deterministic ultrafast all-optical magnetization switching at the nanoscale in such systems, opening a route to develop nanoscale ultrafast magneto-optics.

**MAIN TEXT**

One of the most demanding current technological challenges is the storage and processing of the exponentially increasing amount of information. The quest to improve the density, speed and energy efficiency of magnetic memory storage has led to the exploration of new ways of manipulating magnetism at the ultrafast timescale. A particularly promising possibility was opened up by the discovery of All-Optical Switching (AOS) of magnetization [1] wherein the magnetization of the Rare Earth - Transition Metal (RE-TM) alloy GdFeCo was reversed using ultrashort laser pulses in the absence of an external magnetic field. Following this first observation, AOS has also been found to occur in a variety of materials [2]–[5]. While the performance of AOS in terms of writing speed [6] and energy efficiency [7] compares favorably to other techniques, in terms of the smallest attainable bit size AOS is limited to the micrometer scale owing to the diffraction limit of light. Attempts to downscale the bit size reached down to a few hundred nanometres by tighter focusing of the laser beam using a microscope objective [8] and by nanopatterning [9]–[11]. An important step towards nanoscale bit size was achieved by employing gold two-wire nanoantennas on a TbFeCo alloy thin film, yielding an AOS-switched spot size of 50 nm by exploiting localized plasmons [12]. This suggests that the use of nanoplasmon optics could make AOS technologically viable by making the attainable bit size comparable to that, for example, of heat-assisted magnetic recording [13]. The so far revealed mechanism behind AOS implies that AOS proceeds via fast and efficient demagnetization [6], [14]–[16]. Therefore, the first crucial step towards nanoplasmonic AOS is to follow the temporal evolution of the magnetization in response to plasmon nanoantenna-assisted resonant laser excitation.

Here we uncover the demagnetization dynamics in a ferrimagnetic TbCo alloy nanofilm assisted by nanoring-shaped Ag plasmon antennas. Plasmon nanorings are a widely studied system, displaying two so-called bonding and anti-bonding optical resonances: the anti-bonding (high energy) dipolar mode concentrates the electromagnetic near-field at the rims of the nanoring structure [17], whereas the bonding (low energy) mode does so mostly in the center of the nanoring [17], [18]. The latter has a decisive advantage in the present study as leveraging on the choice between the two modes, since it combines lower energy with smaller mode volume. We compare the dynamics under resonant vs. off-resonant ultrafast laser pumping. In general, we find qualitatively similar sub-picosecond demagnetization for both resonant and off-resonant pumping. However, the degree of demagnetization is substantially smaller for resonant pumping. We attribute this to the strong electromagnetic near-field confinement and the scattered field funneling by the nanoring antenna bonding mode. While the illuminating laser fluence is predominantly channeled to the extremely small areas of the nanorings' near-field, we probe an averaged response largely comprising the signal from the much larger sample areas outside the nanorings, that receive much less fluence and consequently experience a smaller demagnetization. Thus, while an all-optical pump-probe scheme cannot directly yield a complete picture of the nanoscale magnetization dynamics in the near-field, the effects of resonant excitation are clearly detectable. Importantly, through electromagnetic simulations we get further evidence of the funneling of laser fluence right into the centre of the nanoring antennas, supporting the experimental observations.

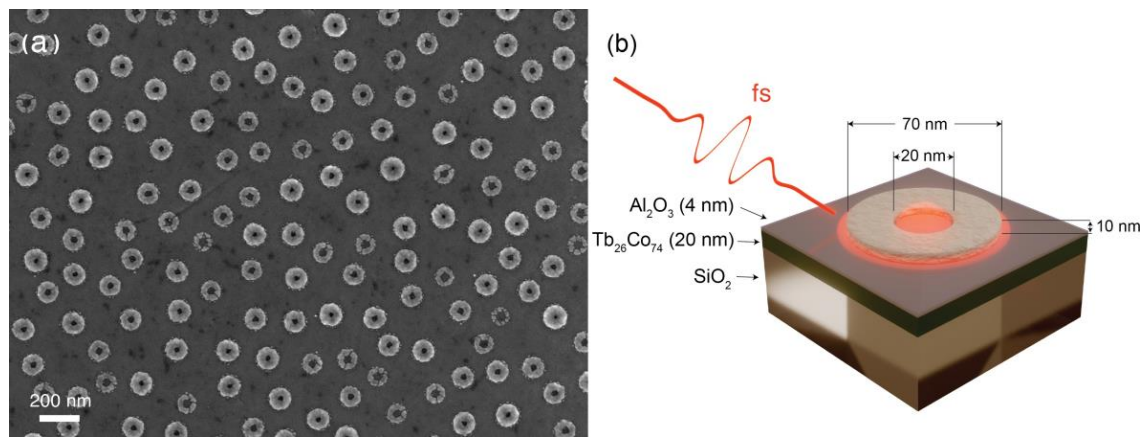

**Figure 1.** (a) SEM overview of the Ag nanoring antennas macroscopic assembly on a $Tb_{26}Co_{74}$ film; (b) A schematic of the nanorings + ferrimagnetic film system (indicating the thicknesses /

dimensions of the elements) and the experimental scheme of exciting the system using an ultrashort laser pulse (antenna near-fields are shown schematically in red).

## Results and Discussion:

### Plasmon nanoring antennas on ferrimagnetic film

Figure 1a shows the scanning electron microscope (SEM) image of the TbCo nanothin film with Ag nanoring antennas directly on top of the few-nm thick sapphire ($Al_2O_3$) capping layer, taking up about 20% of the surface. Nanoantennas are arranged with a short-range order, with the average spacing between the nanorings ensuring the absence of near-field electromagnetic coupling. That is, the spectral response of the entire surface is representative of a single nanoantenna, with the correction of spectral inhomogeneous broadening due to nanoantennas size variations. This is typical for short-range-ordered arrays, produced with hole-mask colloidal lithography, employed here [19]. Laser illumination is done at near-normal incidence (5° off normal), and the film structure underneath the nanoantennas is detailed in Figure 1b.

The optical transmission of the nanoantennas + nanofilm system shows two pronounced resonances (Figure 2a), corresponding to the antibonding (close to 480 nm) and the bonding (close to 920 nm) localized plasmon modes. These are well-studied dipolar modes of the nanoring antennas [17], [18]. Figure 2b shows the calculated charge distribution in the nanoring antennas when each mode is excited. For the bonding (symmetric) mode, the charge distribution has the same sign on the inner and the outer edges of the nanoring, whereas for the antibonding (antisymmetric) mode the charge distribution is the opposite for the inner and the outer edges. Static magnetic characterization (Figure 2c) reveals a square hysteresis loop indicating perpendicular magnetic anisotropy, similar to the hysteresis loops obtained for bare TbCo films [20]. The coercive field increases from 0.2 T for the bare film to 0.26 T for the film with nanoantennas, possibly due to perturbations of the continuous film caused by the antennas' nanofabrication, introducing domain wall pinning sites.

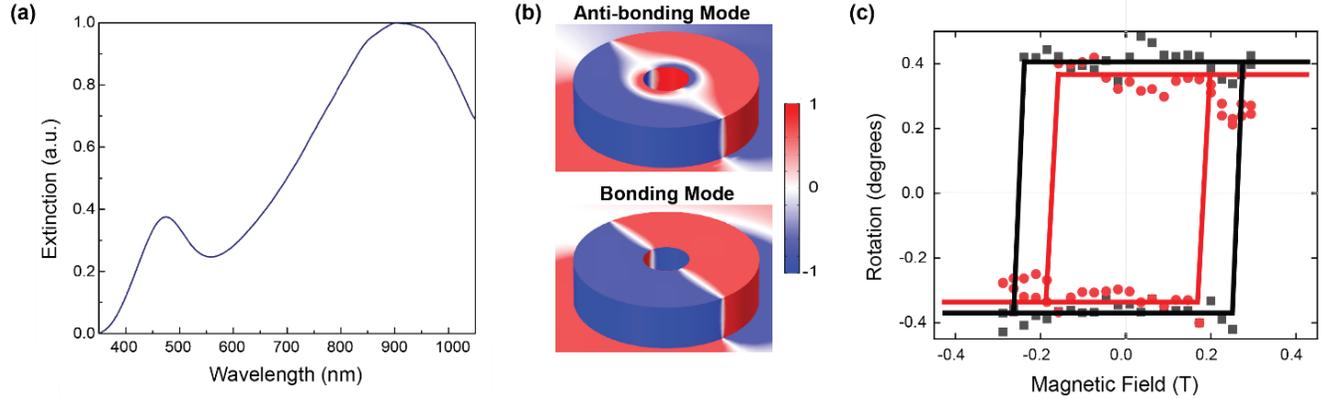

**Figure 2.** (a) Experimentally measured normalized optical extinction spectra showing peaks corresponding to the bonding and antibonding modes; (b) Simulated normalized surface charge distributions corresponding to the two main dipolar plasmon modes of nanoring antennas; (c) Magnetic hysteresis loops for the $Tb_{26}Co_{74}$ film with nanoring antennas (black) and pristine $Tb_{26}Co_{74}$ film (red), showing square hystereses (the solid lines are a guide to the eye).

**Experimental Pump-Probe Dynamics**

The $Tb_{26}Co_{74}$ nanofilm has previously been reported to show multi-shot helicity-dependent AOS [21]. Linearly polarized pump pulse trains or single shots of any polarization induce demagnetization in the film. That is, using a linearly polarized pump beam, we expect to only see the effects of pump-induced heating. The symmetric shape of the nanoring antennas rules out any dependence on the direction of linear polarization of the pump. Thus, we compare the time resolved dynamics of the system for resonant excitation of the dipolar bonding mode (using a pump wavelength of 950 nm), and off-resonant pumping (for a wavelength of 650 nm) with a linearly polarized pump.

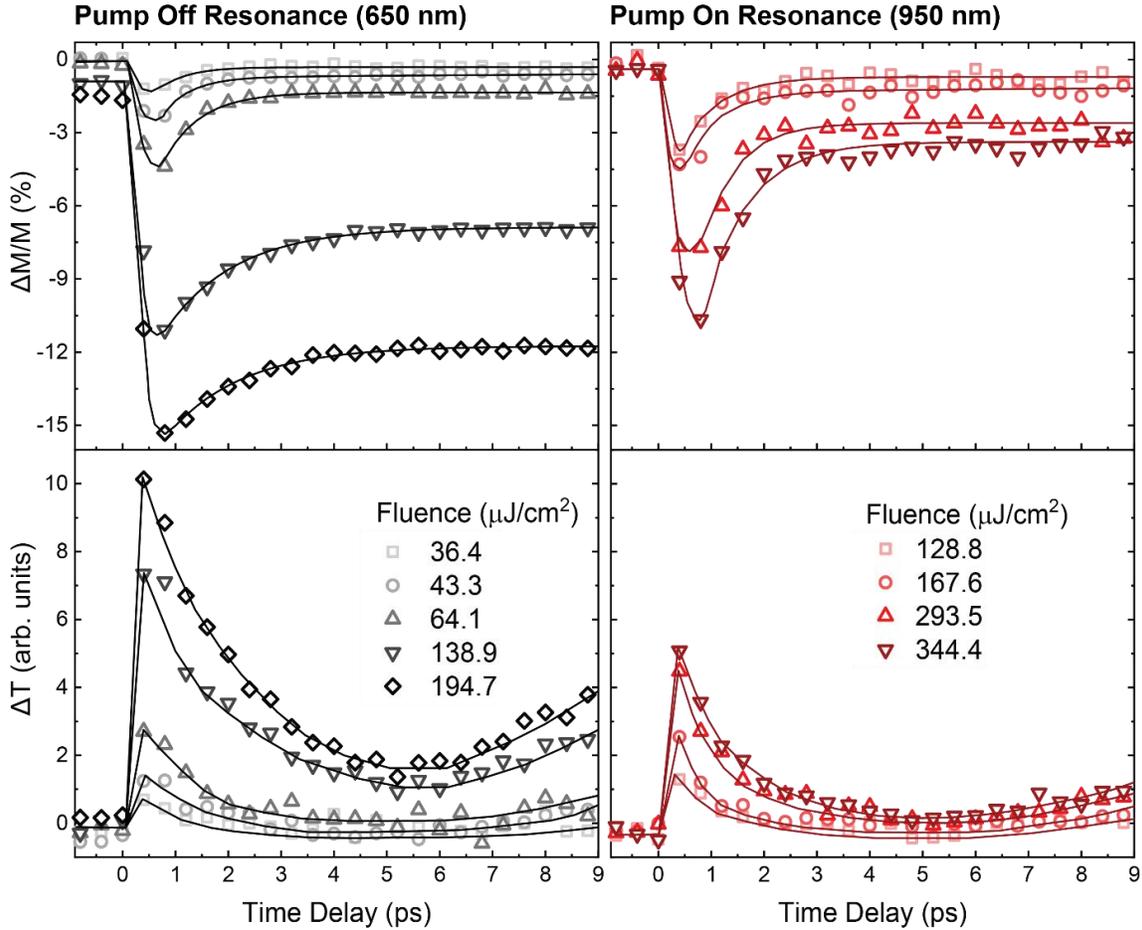

**Figure 3.** Pump-induced demagnetization (top panel) and change in transmission (bottom panels) measured for a range of incident fluences (colors shades code for different fluences) for an off-resonant pump wavelength of 650 nm (left panels, black/grey data points) and a resonant pump wavelength of 950 nm (right panels, red/dark-red points). The solid lines in the case of the magnetization traces are the bi-exponential fitting curves, whereas the solid lines in case of transmission dynamics are a guide to the eye.

The nanoantennas + nanofilm system shows sub-picosecond demagnetization followed by a recovery, similar to the demagnetization behaviour previously reported for $Tb_xCo_{100-x}$ alloys [22]. The nanoantennas do not seem to qualitatively affect the demagnetization process, as evidenced from the similar behaviour for on- and off-resonance excitation (Figure 3). However, the main striking difference between the two cases is the degree of demagnetization. Counterintuitively, larger demagnetization is observed for off-resonance excitation compared to resonant excitation. In Figure 3 the dynamics are plotted for the two cases on the same y-axis to visualize the difference in the degree of demagnetization for selected laser fluences. The same trend is observed for pump-induced changes in transmission.

The difference between the responses for the two cases is even more apparent in Figure 4 where the degree of demagnetization (top panel) and the maximum change in transmission (bottom panel) are plotted as a function of incident pump fluence. A linear relationship is observed for both quantities for both pump wavelengths, as expected for heat-driven dynamics. However, the slope of degree of demagnetization as a function of fluence for the case of off-resonant pumping (0.077) is nearly thrice the slope for resonant pumping (0.028). Similarly, a roughly threefold decrease at resonance is observed for pump-induced transmission changes, where the slope for off-resonant pumping is 0.05, whereas that for resonant pumping is 0.015.

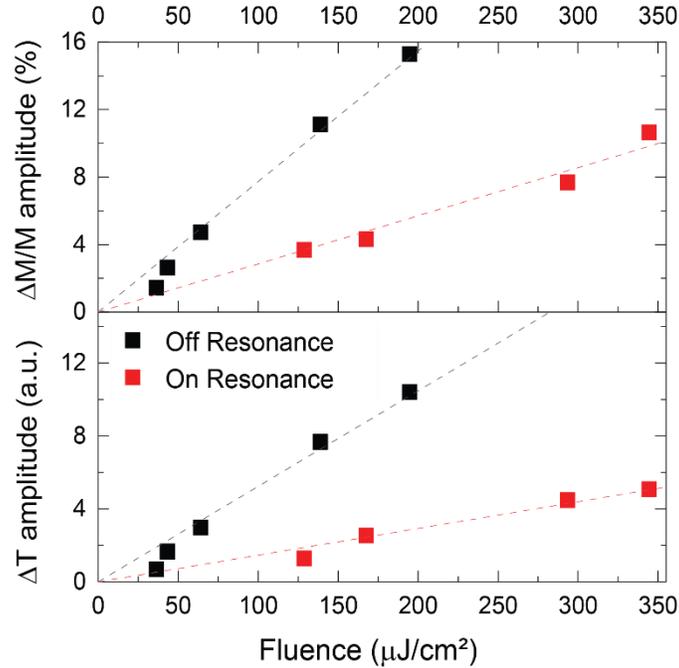

**Figure 4.** Degree of demagnetization (top panel) and maximum change in transmission (bottom panel) plotted as a function of incident fluence for resonant (red squares) and off-resonant pumping (black squares). The dashed lines are the corresponding linear fits to the data points.

The difference, at first sight surprising, in a weaker response for resonant than off-resonant excitation of the system can be rationalized by a funneling of the incident pump fluence to the centre of the nanorings upon exciting the dipolar bonding mode. Indeed, the strongly reduced pump-induced changes observed for resonant pumping can then be explained by considering the size of the probe/illumination spot relative to the size of the nanoantennas. The signal for the μm-sized probe is obtained from an area covering thousands of nanoantennas as well as from the ferrimagnetic film outside of the nanorings' near-field. Crucially, nanoantenna-resonance-enhanced demagnetization is expected to arise only for the TbCo film in the nanorings' focus, where the incoming electromagnetic field of light is funneled and enhanced. However, in our measurements the predominant contribution to the magnetic signal comes from the TbCo outside the nanorings' near-field, receiving considerably less fluence and thus producing less signal.

Knowing the surface coverage of the nanoring antennas on TbCo film allows estimation of the ratio of surface area of the TbCo within the nanoring cavity to the surface area outside of the nanorings, which amounts to 1:44. The predominant contribution to the magnetic signal is thus coming from the TbCo film outside of the nanoantennas near-field. These regions receive considerably less fluence under resonant illumination than under off-resonance illumination due to the efficient in-coupling of the incident laser fluence by the nanoantennas. Note that the nanoantenna optical cross-section is substantially larger than their geometrical size, as it is common for plasmon nanostructures. Though the demagnetization dynamics of the TbCo film in the interior of the nanorings (inner opening of 20 nm) cannot be resolved using an optical probe in our experiments, the signal from areas outside the nanorings' near-field gives an indirect evidence of the electromagnetic field funneling by the plasmon nanoantennas at resonance.

To account for the difference in the system's response at the two different pumping wavelengths, we further take into account the wavelength-dependent absorption of the ferrimagnetic material. Our earlier study [21] provides the optical constants for various compositions of $Tb_xCo_{100-x}$ amorphous alloy films in the wavelength range 400 - 1600 nm. The values of the optical constants as a function of wavelength are very similar for the films with Tb content between 24 - 29 %. As the composition of the films here is in this range, we use these values to extract the absorption at 650 nm and 950 nm (not shown) and find that at a given fluence, the absorption at both wavelengths is identical within 1% uncertainty. That is, the observation of three times smaller pump-induced changes at 950 nm compared to 650 nm strongly supports the hypothesis of plasmon-mediated funneling of the incident fluence.

The effects of plasmon resonance on demagnetization have been previously investigated on a system of gold nanorods on a ferromagnetic permalloy film [23]. This study found enhanced demagnetization of the permalloy at resonance compared to off-resonance excitation, which is opposite to our observations. However, an important difference between this system and the system studied here is the geometry of the plasmon element. Plasmon nanoring antennas are advanced structures in terms of the resonance modes that result from the coupling of inner and outer walls of the nanoring [18][24]. Specifically, the electromagnetic near-field profile of the nanoring antenna features a tightly confined (20 nm in size) and enhanced spot right in the middle of the nanoring, creating the field funneling effect mentioned above. In addition, localized near-fields at the nanoring opening makes this nanoantenna a very prominent candidate for the highly sensitive plasmonic bio- / chemo-detector with an open and easily accessible electromagnetic cavity [25].

**Electromagnetic simulations of the nanoantennas + ferrimagnetic nanofilm system**

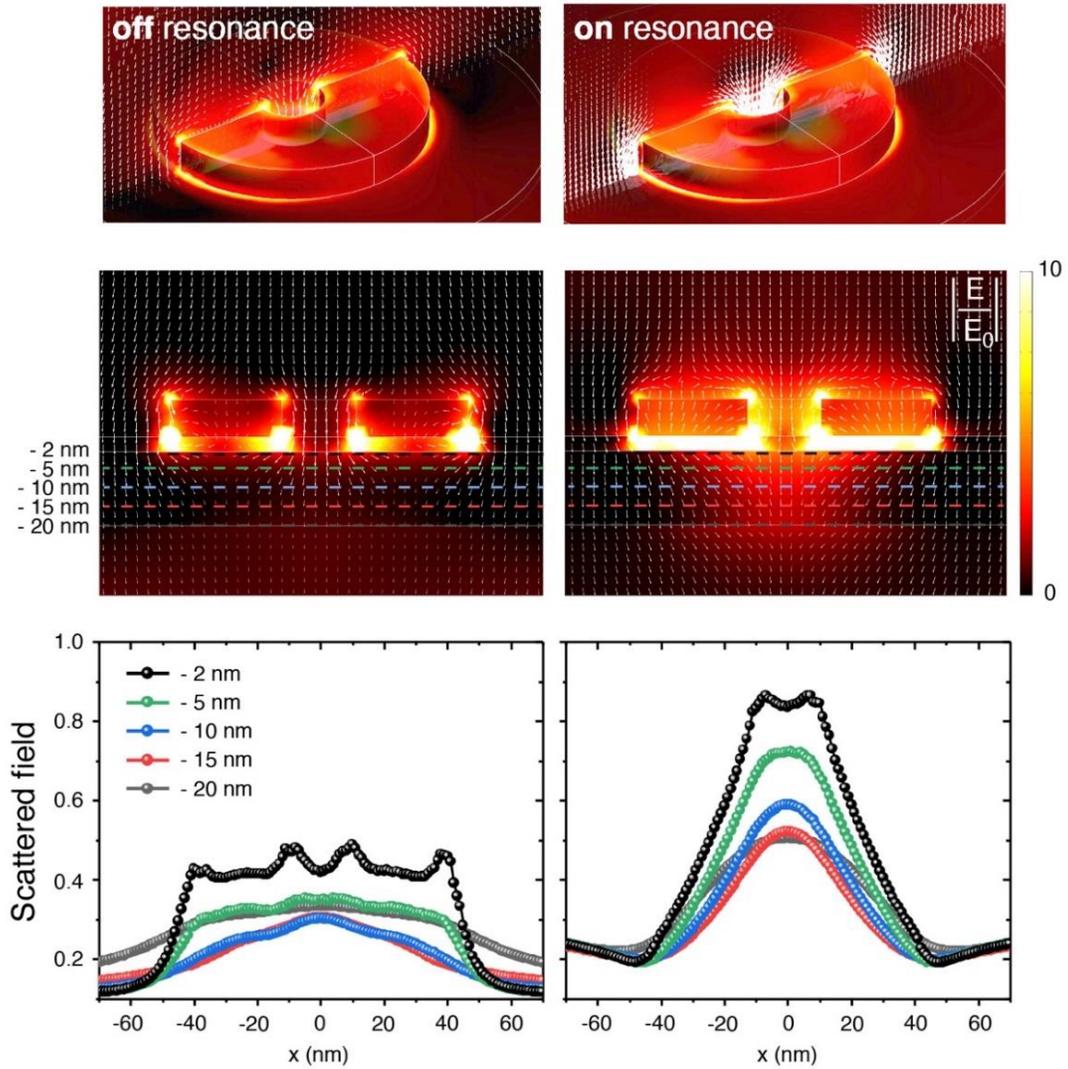

**Figure 5.** (Top) 3D intensity maps showing the scattered electromagnetic field in the x-z plane of the nanoring + ferrimagnetic film for off-resonance (left) and on-resonance (right) illumination. The white arrows represent electric field flux lines. (Middle) Cross-sectional plot of the near-field ($E/E_0$) with dotted differently colored lines marking the distances from the $Al_2O_3$ (capping layer) -TbCo interface. (Bottom) Linear scans of the scattered field along the colored dashed lines of the middle panel.

To support the idea of the field funneling into the TbCo film by nanoring antennas, we performed electromagnetic simulations on nanoantennas + ferrimagnetic film system (Figure 5, off-resonance and on-resonance illumination panels are grouped to the left and right sides, respectively). We visualize the intensity of the scattered field by 3D plots off- and on-resonance (Fig. 5 top), highlighting field funneling by the field flux lines. Cross-sectional intensity maps also show this effect (Fig. 5, mid-panels), along with a substantial field enhancement in the nanoring center for the on-resonance illumination. For the off-resonance case, a strong scattering at the corners is observed, arising from the conventional tip-effect, with the scattered field mostly diffused away.

The strong contrast between off- and on-resonance cases can be appreciated in the magnitude of the scattered field, plotted as a function of position in the $Tb_xCo_{100-x}$ film for an alumina - TbCo interface (Fig. 5, bottom panels). At resonance the nanoring antenna concentrates the scattered field within the nanoring center and directs it towards the underlying substrate of TbCo just below the central opening of the ring. A more defined deflection of the flux lines near the nanoring rims and a strong concentration (about twice the scattering field intensity compared to the off-resonance illumination) at the center of the nanostructure is detected. The funneling effect saturates between 15 nm and 20 nm into the depth of TbCo film, making the thickness of the latter in the present study (20 nm) an exemplary case.

## Conclusion

We find that plasmon nanoring antennas can efficiently funnel the illuminating electromagnetic field into the nanoscopic portions of the ferrimagnetic thin films and induce marked changes in the demagnetization of this system. Upon the resonant excitation, the nanorings are able to couple-in a substantial portion of the incident pump-fluence and concentrate it into the 20 nanometers focal spot of the nanoantenna. This results in smaller pump-induced changes in magnetization and transmission in the areas outside the nanoring antennas that constitute the major part of the studied surfaces that are probed in the macroscale pump-probe experiments. Electromagnetic simulations of the scattered field show qualitative agreement with this picture.

Overall, studying the nanoscale confinement of demagnetization processes under the influence of plasmon nanoantennas fundamentally require experiments with higher real-space resolving power. These are extremely experimentally demanding and are often requiring large-scale experimental facilities (such as the measurements of the results of the fs-pulsed illumination in ferrimagnets with photoelectron emission microscopy, PEEM [10] or with X-Ray Holographic Imaging combined with X-Ray Magnetic Circular Dichroism [12]). Here we essentially circumvent the need for such demanding experiments by rationalizing the fundamental role of nanoantennas in focusing the pulsed-light illumination to the nanoscale. A further step in this direction is to reduce the ferrimagnetic film to nanoscopic elements positioned in focus of plasmon nanoantennas while strictly maintaining their magneto-optical properties and required magnetic anisotropy [26]. Such an experiment would provide, conversely, nanoantenna-enhanced demagnetization.

In the present case the proposed incident fluence funneling to the nanoscopic regions of the ferrimagnet marks a promising path towards ultrafast magnetic bit miniaturization even for the nanofilm systems, broadly currently employed. We envision such a path in practice leading to fully functional ultrafast nanoscale magnetic memory architectures.

## Methods

The sample investigated consists of a $Tb_{26}Co_{74}$ amorphous alloy thin film grown on a glass substrate, covered with an alumina capping layer on top of which plasmonic silver nanorings have been fabricated.

The 20 nm thick TbCo film was prepared by DC magnetron sputtering from elemental Tb and Co targets. To ensure uniform film deposition at room temperature, a rotating sample holder was used. The synthesis was performed under ultrahigh vacuum, with a base pressure of $10^{-10}$ Torr and an $Ar^+$ sputtering gas pressure of 2-3 mTorr. The TbCo film was covered with a 4nm $Al_2O_3$ capping layer. More details of the synthesis can be found in [20] and [21]. Silver nanorings of inner diameter 20 nm, outer diameter 70 nm and height 10 nm were fabricated on the capped TbCo film by hole-mask colloidal lithography, HCL [19]. Fig.1(b) shows the structural details of the sample.

The fabricated sample was imaged using a Scanning Electron Microscope (SEM). To characterize the resonance modes for the sample, the optical transmission spectrum was measured in the wavelength range 350-1050 nm. Static magneto-optical characterization was done using a polar Faraday geometry at 800 nm.

Magnetization dynamics were measured using an all-optical two-colour pump-probe setup. The probe was derived from a Ti:sapphire amplified laser system with a 1 kHz repetition rate, a central wavelength of 800 nm, and a pulse width of 100 fs at the sample position, focused to a spot size of 460 µm. The pump pulse was derived from the same laser by tuning the wavelength through optical parametric amplification. The off-resonance pump wavelength was chosen to be 650 nm, and the spot size at the sample was 510 µm. The resonant pump wavelength was chosen as 950 nm, and the spot size at the sample was 590 μm. The setup was built to have near normal incidence of the pump (~ 5° to the sample normal). Both pump and probe beam were horizontally polarized i.e., in the plane of incidence. The probe beam was separated from the pump using the appropriate colour filters and was detected as a function of the pump-probe time delay using a pair of balanced Si photodiodes. A static magnetic field higher than the sample coercive field was applied normal to the sample throughout the course of each measurement to re-initialize the saturated magnetic state of the sample before each subsequent pump pulse.

Three-dimensional electrodynamic calculations of the optical response and the surface charge density maps were performed by solving the Maxwell equations via the finite element method (FEM) implemented in the commercial COMSOL Multiphysics software [27]. In order to reproduce the experimental structures, we modeled a silver ring on a three-layered structure using an analytical background field resulting from solving Fresnel equations for an incoming light source on a multilayer system. After the interaction with light, the scattered field due to the silver ring was calculated.

**ACKNOWLEDGMENT**

The authors thank Dr. Oleg Lysenko for his contribution to samples nanofabrication. K.M, A.V.K and A.K thank Dr. Sergey Semin and Chris Berkhout for technical support. This work was supported by the project FEMTOTERABYTE funded from European Union's Horizon 2020 research and innovation program under grant agreement no. 737093. PV acknowledges support from the Spanish Ministry of Science and Innovation under the Maria de Maeztu Units of Excellence Programme (MDM-2016-0618), and the project RTI2018-094881-B-I00 (MICINN/FEDER). V.K. acknowledges support from the Swedish Research Council (Project No. 2019-03581).